\documentclass[12pt]{book}

\usepackage[dvips]{graphicx,color}
\usepackage{makeidx,tsukuba}

\makeauthorindex
\makeindex

\begin{document}

\BookTitle{\itshape The 28th International Cosmic Ray Conference}
\CopyRight{\copyright 2003 by Universal Academy Press, Inc.}
\pagenumbering{arabic}

\chapter{ Hybrid Performance of the Pierre Auger Observatory and Reconstruction of Hybrid Events }

\author{
  B. Fick,$^1$  for the Pierre Auger Collaboration,$^2$\\
{\it (1) The Enrico Fermi Inst., the University of Chicago, 5640 S. Ellis Av., Chicago, Illinois, USA\\
(2) Observatorio Pierre Auger, Av.San Martin Norte 304, 5613 Malargue, Argentina} \\
}

\section*{Abstract}
The Pierre Auger Observatory is a``hybrid'' UHECR detector.
The surface detector (SD) and air fluorescence detector (FD) of
the observatory are designed for observation of cosmic ray showers in
coincidence, with a 10\% duty cycle. The resulting data are expected 
to be superior in quality to those of either the SD or FD operating 
individually. Hybrid operation, triggering, data acquisition, and event
reconstruction were successfully demonstrated during the prototype phase 
of the project. This paper focuses on 75 hybrid events recorded 
during a four month period of running with the prototype detectors in 
late 2001 and early 2002. The geometric technique for reconstruction 
of the hybrid events is described and its advantages over the traditional
FD-only method are demonstrated. A lateral distribution for the water Cherenkov signals, derived from these data, is presented. 

\section{Introduction}

During the hybrid operating period the Auger engineering prototype observatory consisted of an array of 30 SD stations positioned on a triangular grid of 1.5 km spacing. Two prototype
FD telescopes, located some 10 km away, viewed the volume of air above the 
array. The prototype was constructed to test the feasibility of the
 initial design and to discover any changes required for the complete
 observatory. The hybrid prototype has enabled us to test new concepts in triggering, 
data communications, merging, and reconstruction procedures.
In particular we were able to establish the feasibility of the
hybrid fitting scheme from an investigation of the data. Additionally, we were able 
to check to see whether we can construct a sensible average lateral distribution from 
the combination of FD and SD information. 

\section{Hybrid Operation}  
The hybrid system was built with a ``cross-triggering'' capability.
Data was recovered from both the FD and SD whenever either system 
was triggered. The SD data recorded as a result of an FD trigger
was tagged accordingly. Hybrid events were built on the basis of this
tag or time matching. In most cases the FD, having the 
lower energy threshold, promoted a subthreshold SD trigger.  This is 
an important capability. It will be shown in the next section that timing 
information from even one SD station can improve the geometric reconstruction
of a shower. The primary effect of this was to bring the hybrid energy
threshold well below the design threshold of 10 EeV. 

The hybrid prototype was operated on every available clear moonless
night during the months of December 2001, January 2002, February 2002, and
March 2002. At the time of these measurements neither of the detectors was
satisfactorily calibrated but development changes on either instrument were
held to a minimum during this period. 75 hybrid events were successfully
recorded at a rate of approximately one every 1.5 hours. A number of potential hybrid events were lost during operation
 because of network and clock synchronization problems.

\section{Hybrid Geometric Reconstruction}
 FD shower axis reconstruction proceeds in two stages.
 First, the shower detector plane (SDP) is derived from the angular pattern of hit FD pixels.
 The SDP is the plane containing 
the shower axis and the FD. Second, the shower spot angular motion is 
used to determine the orientation and location of the shower axis in the
SDP. The axis is most conveniently described in terms of the SDP normal
$\hat{N}_{pl}$, the perpendicular distance from FD to shower axis $R_{p}$, the angle
$\chi_{0}$ that the axis make with the horizontal, in the SDP, and $T_{0}$ the 
time at which the shower passes closest to the FD. We call this the ``mono'' geometric
 reconstruction method because it relies on information from a single FD telescope.  
The details of this procedure are discussed elsewhere [1].

 The mono procedure suffers from having to find three geometric parameters
$(R_{p}, \chi_{0},$ and $T_{0})$ from a nearly linear relationship between spot 
position and time. The Auger hybrid procedure solves this problem by 
exploiting the time of shower front arrival at one or more SD stations. 
For any geometry specified by $(R_{p}, \chi_{0},$ and $T_{0})$, there is an 
an expected time when the shower front passes any position in space. The times recorded by 
SD stations are used to help pick out the best geometry by requiring
that their times agree with the expected times at their locations.

We have done a bootstrap[2] analysis of a hybrid event to 
illustrate the situation in Figure 1. On the left we plot 1000 bootstrap
 core location solutions for mono and hybrid fits. On the right we
plot the corresponding $(R_{p},\chi_{0})$ solutions. The mono cores are
 spread out 
along the line of the SDP whereas the hybrid cores are well contained in a smaller
region with greatest extent in the direction perpendicular to the line of the SDP. The
expected ambiguity in the mono determination of $R_{p}$ and $\chi_{0}$ is
illustrated
in the right plot. There is a strong correlation between these two
parameters. The hybrid solutions are more tightly confined and
show no correlation between the parameters. The hybrid axis is much better
determined.

\begin{figure}[t]
  \begin{center}
   \includegraphics[width=13cm,height=7cm]{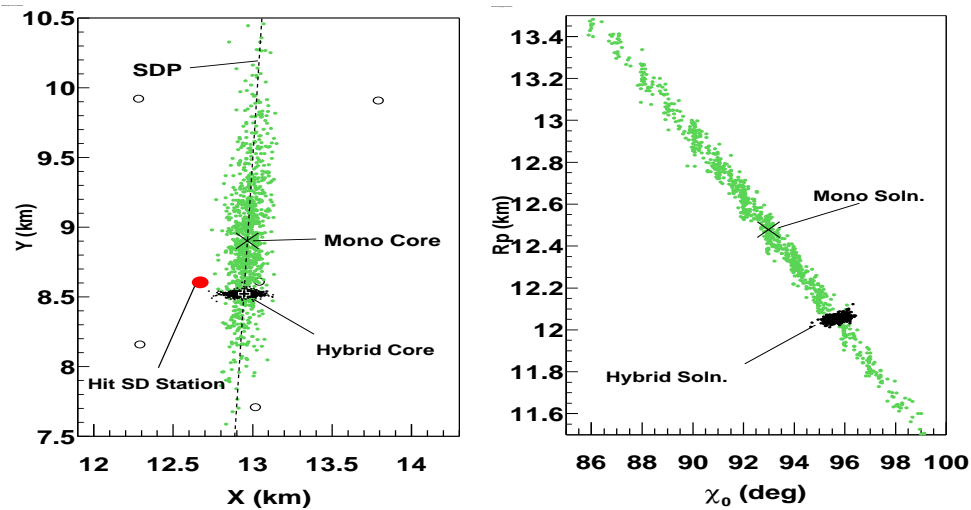}
  \end{center}
  \vspace{-0.5pc}
  \caption{Bootstrap solutions for mono and hybrid fitting procedures for a selected hybrid event.}
\end{figure}

We have done a bootstrap analysis of a subset of the hybrid data set. The data consisted of 38 events
with estimated energies above 1 EeV.  Please refer to [3] for an explanation of FD energy reconstruction. But for the energy cut these events represent a random sample of hybrid triggers that are not optimized for favorable mono fitting. 
This is shown in Table 1. The hybrid procedure determines geometries with greater precision. 
\begin{table}[ht]
 \caption{Bootstrap uncertainties for geometric reconstruction.}
\begin{center}
\begin{tabular}{l|ccccc}
\hline
 Method  & $\langle\delta R \rangle$ (m) &      Max $\delta R$ (m)       & $\langle\delta\chi_{0} \rangle$ $(^{\circ})$     & Max $\delta\chi_{0}$ $(^{\circ})$ \\
\hline
Mono       & 921     & 2600 &   8.05       & 24     \\
Hybrid    & 20.95 & 86     & 0.24     & 0.7   \\
\hline
\end{tabular}
\end{center}
\end{table}
To investigate the sensitivity to poor timing we introduced timing errors 
into the hybrid reconstruction chain and determined the effects on geometry for 68  hybrid 
showers. The mean change of FD core distance $R$ per 100 ns clock offset
was 39 m. 90\% of the events changed by less than 70 m. 
The mean change in $\chi_{0}$ was 0.26$^{\circ}$. 90\% of the events changed by less than 0.5$^{\circ}$. 
We conclude that even with a clock accuracy of 100 ns we will get 
geometric reconstructions exceeding our original design specification.

\section{Combined LDF}
We have combined information from both the FD 
and SD to form an average lateral distribution. The geometry of each
event was found using the hybrid fitting method described above. Events with 
as few as one triggered SD stations are included.   
Information from SD stations were 
included based on the following criteria : $\delta R < 100 m$ , $\delta X_{core} < 100 m $, $\delta E / E < 0.2$, and $\theta_{} < 60^{\circ}$ . For each 
qualifying SD station we plot , in Figure 2, the observed particle density normalized by the FD energy 
versus its  distance to the core. Horizontal Error bars come from a bootstrap analysis of
the hybrid-derived geomety . Vertical error bars are a combination of uncertainies in SD calibration and FD 
energy. A curve representing a power law form for the lateral distribution function, $S(r) = k r^{-\nu}$ with $\nu = 2.99 $ has been drawn through the points. The agreement in shape between data and model is apparent.    
\begin{figure}[htb]
  \begin{center}
    \includegraphics[width=12cm]{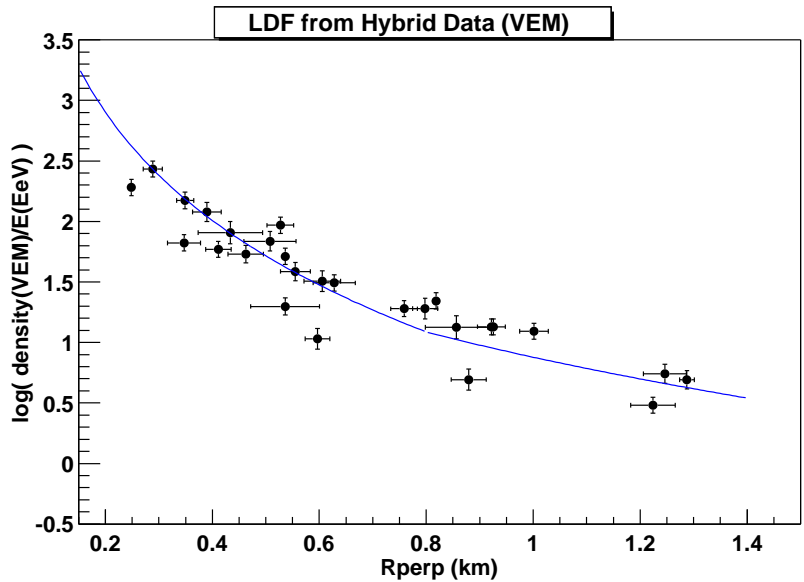}
  \end{center}
  \vspace{-0.5pc}
  \caption{Average lateral distribution of normalized sd station densities.}
\end{figure}

\vspace{\baselineskip}

\re
1.\ P.Privitera for the Pierre Auger Observatory, these proceedings
\re
2.\ Efron,B., et. al., ``An Introduction to the Bootstrap'', Chapman and Hall(1986).
\re
3.\ S.Argir\`{o} for the Pierre Auger Observatory, these proceedings
\re

\endofpaper
\end{document}